\documentclass[pdflatex,sn-mathphys-num]{sn-jnl}
\usepackage[T1]{fontenc}%
\usepackage{graphicx}%
\usepackage{multirow}%
\usepackage{amsmath,amssymb,amsfonts}%
\usepackage{amsthm}%
\usepackage{mathrsfs}%
\usepackage[title]{appendix}%
\usepackage{xcolor}%
\usepackage{textcomp}%
\usepackage{manyfoot}%
\usepackage{booktabs}%
\usepackage{array}%
\usepackage{rotating}%
\usepackage{caption}%
\usepackage{threeparttable}%
\usepackage{url}
\usepackage{hyperref}
\theoremstyle{thmstyleone}%
\theoremstyle{thmstyletwo}%
\theoremstyle{thmstylethree}%

\begin{document}

\title{Auditing Reproducibility in Non-Targeted Analysis: 103 LC/GC--HRMS Tools Reveal Temporal Divergence Between Openness and Operability}

\author*[1,2]{\fnm{Sarah} \sur{Alsubaie}} \email{sarah.alsubaie@kaust.edu.sa}
\author[2,3]{\fnm{Sakhaa} \sur{Alsaedi}}\email{Sakhaa.Alsaedi@kaust.edu.sa}
\author[1,2,3]{\fnm{Xin} \sur{Gao}}\email{Xin.Gao@kaust.edu.sa}

\affil*[1]{\orgdiv{Bioscience Program, Biological and Environmental Science and Engineering}, \orgname{King Abdullah University of Science and Technology (KAUST)}, \orgaddress{\street{4700}, \city{Thuwal}, \postcode{23955--6900}, \state{Mecca Region}, \country{Kingdom of Saudi Arabia}}}
\affil[2]{\orgdiv{Center of Excellence for Smart Health (KCSH) and, Center of Excellence on Generative AI}, \orgname{King Abdullah University of Science and Technology (KAUST)}, \orgaddress{\street{4700}, \city{Thuwal}, \postcode{23955--6900}, \state{Mecca Region}, \country{Kingdom of Saudi Arabia}}}
\affil[3]{\orgdiv{Computer Science Program, Computer, Electrical and Mathematical Sciences and Engineering Division}, \orgname{King Abdullah University of Science and Technology (KAUST)}, \orgaddress{\street{4700}, \city{Thuwal}, \postcode{23955--6900}, \state{Mecca Region}, \country{Kingdom of Saudi Arabia}}}

\abstract{
In 2008, melamine in infant formula forced laboratories across three continents to verify a compound they had never monitored. Sudan dyes in spices and nitrosamines in pharmaceuticals created similar emergencies. Non-targeted analysis using LC/GC–HRMS handles these cases. But when findings trigger regulatory action, reproducibility becomes operational: can an independent laboratory repeat the analysis and reach the same conclusion?
We assessed 103 tools (2004–2025) against six pillars drawn from FAIR and BP4NTA principles: laboratory validation (C1), data availability (C2), code availability (C3), standardised formats (C4), knowledge integration (C5), and portable implementation (C6). Health contributed 51 tools, Pharma 31, and Chemistry 21.
Nine in ten tools shared data (C2, 90/103, 87.38\%). Fewer than four in ten supported portable implementations (C6, 40/103, 38.83\%). Validation and portability rarely appeared together (C1×C6, 18/103, 17.48\%). Over twenty-one years, openness climbed from 56\% to 86\% while operability dropped from 55\% to 43\%, the gap widening from near parity to 43.7 percentage points. No pairwise association survived Bonferroni correction. No tool addressed food safety.
Journal data-sharing policies achieved one goal and missed another: they increased what authors share but not what reviewers can run. Tools became easier to find but harder to execute. Food safety requires validated, standardised, and portable workflows—the exact combination of current tools least provide. Strengthening C1, C4, and C6 would turn documented artifacts into workflows that external laboratories can replay.
\textbf{Scientific Contribution}
We audited 103 LC/GC–HRMS tools for reproducibility across Health, Pharma, and Chemistry—the first cross-domain architectural assessment at this scale. Previous studies examined analytical performance; we examined whether an external laboratory could rerun the workflow. The framework tells developers what to check: not just whether code is shared, but whether it runs.
}

\maketitle

\section{Background}\label{sec:background}

\begin{figure}[htbp]
\centering
\includegraphics[width=0.85\textwidth]{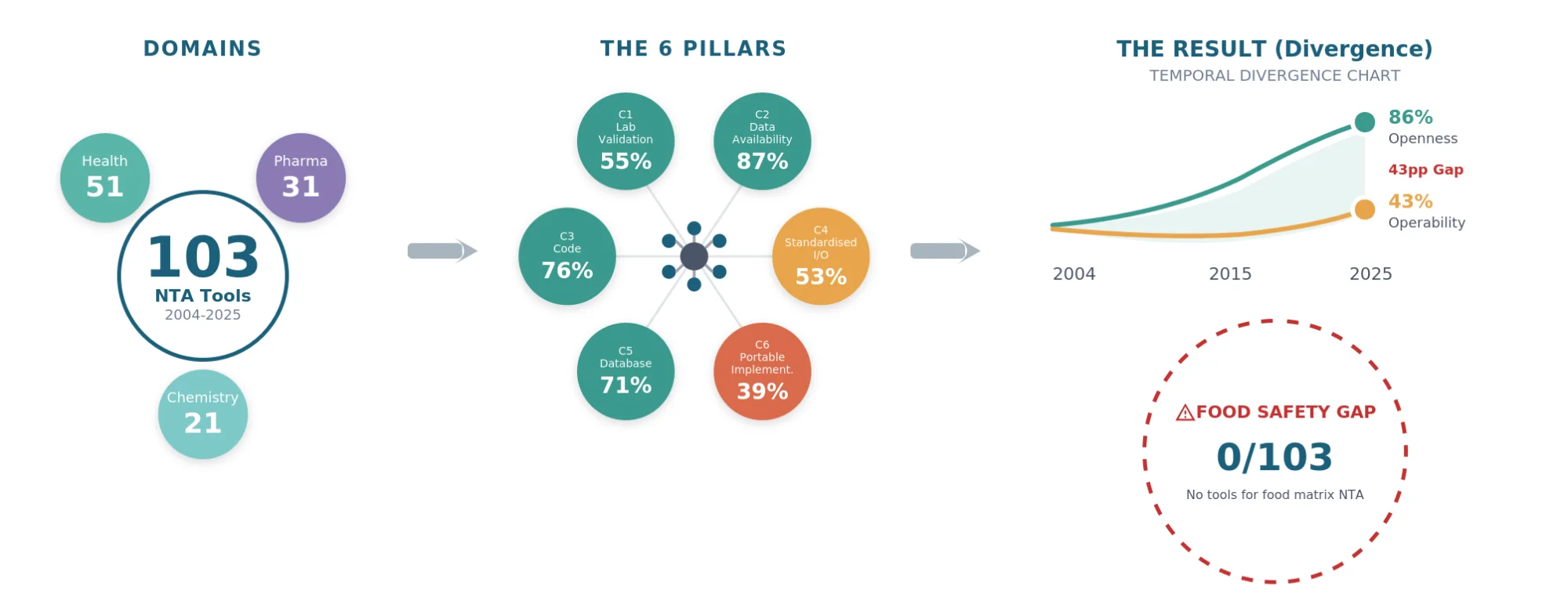}
\caption{Temporal divergence between openness and operability in NTA tools (2004--2025). Journal mandates increased sharing (openness pillars: C2, C3, C5) but not functionality (operability pillars: C1, C6).}
\label{fig:Abstract}
\end{figure}

In 2008, melamine contamination in Chinese infant formula forced laboratories across three continents to detect a compound they had never monitored \cite{pei2011,wu2013analytical,mazaheri2024method}. Similar incidents emerged repeatedly. Between 1990 and 2009, diethylene glycol in paediatric paracetamol caused kidney failure outbreaks \cite{soleman2024effects}. Industrial dyes such as Sudan Red appeared in spices \cite{rebane2010review,cornet2006development}. Nitrosamine impurities emerged in widely prescribed pharmaceuticals \cite{bhirud2024nitrosamine,ruepp2021eu}. Per- and polyfluoroalkyl substances (PFAS) reached drinking water supplies \cite{megson2024non,sadia2023occurrence}. In each case, authorities had to act on compounds that were never on any predefined target list \cite{barzen2017discovery,ting2006role}. The analytical identification became the basis for product recalls, import detentions and public-health advisories Figure~\ref {fig:Abstract}. It had to withstand regulatory audit and legal challenge \cite{ghijs2024continuing}.

These incidents share one problem: laboratories must identify compounds they never expected to find, often within days, and their findings must hold up across borders. When findings trigger regulatory action, reproducibility shifts from scientific ideal to operational requirement. The question becomes whether an independent laboratory can verify the claim without privileged access to the originating site. NTA platforms must deliver architectural guarantees, not 
just analytical performance. Regulatory compliance and public health depend on it. Reproducibility must be built into the infrastructure, not retrofitted after findings emerge. Understanding why requires distinguishing targeted from non-targeted approaches.

Classical targeted analysis works when the contaminant is known and available as a reference standard \cite{amaral2021target}. It fails when the species is novel, intentionally disguised or absent from surveillance lists. This gap has driven the rise of NTA. LC/GC--HRMS acquires full-scan data across broad mass ranges, prioritises unexpected features and retrospectively interrogates stored data for compounds not anticipated at acquisition \cite{Hollender2017NTS,simonnet2022evidence}. LC–HRMS targets polar and thermolabile compounds; GC–HRMS serves volatile and semi-volatile species; both share the non-targeted acquisition logic. LC/GC--HRMS NTA platforms now inform decisions in pharmaceutical quality control, clinical diagnostics and environmental monitoring precisely because action must be taken on emerging or previously unrecognised chemical signals \cite{efsa2021,lai2021retrospective}.

When NTA findings trigger regulatory action, data and scripts deposited in repositories are necessary but not sufficient \cite{Wilkinson2016,Sandve2013TenSimpleRules}. It must mean that an independent group, potentially in a different jurisdiction running on a different compute stack, can begin from the declared inputs, execute the declared workflow and reach concordant conclusions within agreed tolerances \cite{Peng2011,baker20161}. The question shifts from "did you find something?" to "can someone else verify what you found?" \cite{goodman2016reproducibility}. Reproducibility becomes a property of the analytical architecture \cite{ng2020assessing,nab2024opensafely}.

That property is determined by design decisions that either enable portability or trap findings at the site where they were produced \cite{tsugawa2011practical,siskos2017interlaboratory,yang2018empirical}. 
Open file formats reduce vendor lock-in: standards such as mzML and mzTab allow different tools to read the same data, while stable identifiers such as InChIKey ensure consistent molecular reference \cite{Heller2015}. 

Provenance frameworks record analytical lineage. W3C PROV ontology and packaging conventions such as Research Object Crate (RO-Crate) document how data, parameters and software environments produce a reported call \cite{belhajjame2013prov,ROCrate2021}. Workflow engines such as Nextflow and Snakemake decouple pipeline logic from execution environment; Galaxy provides a web-based layer with similar portability guarantees \cite{di2017nextflow,Koster2012Snakemake,afgan2018galaxy,Wratten2021}. 

These systems treat workflows as governed artefacts \cite{colombo2020value}. Each design choice determines whether an external assessor can reconstruct the path from raw signal to reported call without privileged access \cite{namli2024scalable,geisler2021knowledge}.

Auditing current practice reveals what is already hardened and what remains aspirational. This study treats LC/GC--HRMS NTA platforms as cheminformatics infrastructure and asks which architectural guarantees are already in place. We assess six pillars---validation, data availability, algorithmic transparency, standardised formats, knowledge integration and portable implementation---across Health, Pharma and Chemistry domains~\cite{rao2018validation,butin2016formal,zhang2025prioritizing,Martens2011mzML,Griss2014mzTab,Kanehisa2000KEGG,Wishart2018DrugBank,wenteler2024ai,bettanti2024exploring,di2017nextflow,Koster2012Snakemake,afgan2018galaxy,Wratten2021}. These pillars collectively express regulatory-grade reproducibility: a level of reproducibility that can withstand independent audit, reuse and legal scrutiny across organisational boundaries.

Food safety represents a critical yet underserved application area. Incidents such as melamine in infant formula and Sudan dyes in spices required rapid cross-border verification---exactly the scenario demanding portable, validated NTA workflows. Whether current tools address this need---or whether the field's architectural trajectory actively precludes food safety applications---remains unexamined. This study provides that examination.

These considerations motivate three questions: (1)~What is the current adoption rate of each pillar across NTA tools? (2)~Do patterns differ by domain---health, pharmaceutical quality, or chemistry? (3)~Which architectural gaps explain the absence of food safety applications?

A scoping search of PubMed and Google Scholar (October 2025) using terms 'NTA reproducibility audit' and 'non-targeted analysis benchmark' returned no cross-domain architectural assessments at comparable scale. To our knowledge, this represents the first such audit \cite{place2021introduction,baker20161,Peng2011}. Previous studies addressed analytical performance, spectral quality and annotation confidence \cite{hulleman2023critical,lennon2024harmonized}. Whether workflows can be replayed across institutions received less attention. The Benchmarking and Publications for Non-Targeted Analysis Working Group (BP4NTA) established minimum reporting standards; the Findable, Accessible, Interoperable, Reusable (FAIR) principles defined desirable properties for research data \cite{place2021introduction,Wilkinson2016}. Both prescribe what should be reported. They do not check whether current tools actually deliver these properties. Without empirical assessment, we cannot distinguish what is already in place from what remains aspirational. This study addresses that gap \cite{kanwal2017investigating}.

Since approximately 2015, major journals have mandated data and code availability statements as conditions of publication. Repositories such as GitHub, Zenodo and Figshare matured in parallel, making deposition technically straightforward. These policies demonstrably increased sharing rates. What remains unquantified is whether increased availability translated into increased usability---whether tools that are now easier to find are also easier to run.

The intention is not to endorse any specific software. It is to establish a defensible baseline that can be interrogated, reused and challenged. Section~\ref{sec:methods} operationalises the six pillars and explains how tools were assessed. Sections~\ref{sec:results} and \ref{sec:discussion} present adoption patterns and interpret their implications for food safety and environmental monitoring. Section~\ref{sec:conclusions} establishes this audit as a reusable baseline for future platform development.

\section{Methods}\label{sec:methods}

\subsection{Study design}

This study is an architectural audit of LC/GC--HRMS NTA tools. Rather than synthesising outcomes, it maps structural properties against predefined reproducibility criteria.

NTA workflows increasingly inform regulated decisions in pharmaceutical quality, clinical diagnostics, and environmental monitoring. Food safety represents an emerging application area. In such settings, an annotation functions as evidence a third party must be able to rerun the workflow and reach a concordant conclusion. We call this \textit{regulatory-grade reproducibility}: demonstrable capacity for an independent laboratory to replicate findings within predefined acceptance criteria, using documented procedures.

\subsection{Tool identification}

We searched Google Scholar from September 2024 through October 2025 using: ("non-targeted analysis" OR "nontargeted analysis" OR "untargeted analysis" OR "suspect screening") AND ("LC-MS" OR "GC-MS" OR "LC-HRMS" OR "GC-HRMS") AND ("tool" OR "software" OR "platform" OR "workflow"). Monthly alerts captured new publications. We also searched GitHub, GitLab, and Zenodo for tools with limited journal coverage. We applied no domain restriction; health, pharmaceutical, chemistry, food safety, and environmental applications were all within scope. This search window (13 months) captured tools published between 2004 and 2025, as older tools often appear in recent reviews or benchmarks. Table S1 contains platform metadata, search strings, keyword lists, and scoring formulas.

Inclusion required NTA, documentation sufficient to assess at least one pillar, and public accessibility. We excluded tools limited to targeted analysis or lacking implementation. Initial searches yielded 247 candidate tools. After removing duplicates and applying inclusion criteria, 156 remained for assessment. We excluded 53 for insufficient documentation, yielding a final corpus of 103 tools.

The final corpus comprised 103 tools (Table S1), assigned to Health (51/103, 49.51\%), Pharma (31/103, 30.10\%), or Chemistry (21/103, 20.39\%) based on documented application (Table S3). When documentation referenced multiple domains (n=17), we assigned tools using a hierarchy: first, the developer's stated target application; second, the journal venue and described use case; third, the example datasets provided. The first applicable criterion determined assignment. Each tool was then assessed against six reproducibility pillars.

\subsection{Pillar definitions}

Six pillars operationalise reproducibility as externally inspectable signals (Table~\ref{tab:pillars}). C1 addresses empirical practice; C2–C6 address computational infrastructure. The framework synthesises FAIR principles \cite{Wilkinson2016} and BP4NTA reporting standards \cite{peter2021nontargeted}. Provenance 
and version control, sometimes proposed as separate criteria, are captured here within C3 (code repositories track versions) and C6 (workflow engines emit provenance).

\begin{table}[ht]
\centering
\caption{Reproducibility pillars and scoring criteria.}
\label{tab:pillars}
\begin{tabular}{llp{7cm}}
\toprule
Code & Pillar & Criterion \\
\midrule
C1 & Laboratory validation & Documented verification with performance metrics \\
C2 & Data availability & Dataset accessible via URL, DOI, or repository \\
C3 & Code availability & Source code in repository or package index \\
C4 & Standardised I/O & Open vendor-neutral formats (mzML input; mzTab, CSV, or JSON output) \\
C5 & Knowledge integration & Integration with curated databases (e.g., PubChem, HMDB, MassBank) \\
C6 & Portable implementation & Workflow engine (Nextflow, Snakemake, CWL) and/or containerisation (Docker, Singularity) with documented execution \\
\bottomrule
\end{tabular}
\end{table}

We chose binary scoring (present/absent) over ordinal scales to maximise 
reproducibility: a future auditor applying the same criteria should reach 
identical scores.

\paragraph{Conceptual grouping.} For analytical purposes, we distinguish \emph{openness pillars} (C2 data availability, C3 code availability, C5 database integration) from \emph{operability pillars} (C1 validation, C6 portable implementation). C4 (standardised I/O) bridges both groups: standardised formats serve accessibility (enabling data exchange) and executability (enabling tool interoperation) equally, so assigning C4 to either composite would be arbitrary. This distinction enables assessment of whether sharing practices and execution practices advance in parallel.

\subsection{Assessment}

We scored pillars as binary: present (1) when explicit evidence appeared in peer-reviewed text or official project documentation (GitHub repositories, package documentation, developer-maintained websites); absent (0) otherwise.

We read each publication in full and we extracted relevant metadata into structured fields. For C1 (laboratory validation), a score of 1 required experimental verification on real samples or reference standards, with quantified performance. For analytical tools: accuracy, precision, recovery, or detection limits. For ML-based annotation modules: recall, F1, or AUC. Purely computational validation without experimental confirmation scored 0.

For C2–C6, we used Excel's SEARCH function to detect predefined keywords in extracted metadata. We applied SEARCH case-insensitively; we consolidated variant spellings into canonical terms before scoring. Negative modifiers ('not available', 'discontinued') were checked manually; presence of such modifiers overrode keyword detection. For example, C2 scored 1 when repository identifiers (Zenodo, GitHub, Figshare) appeared, and 0 otherwise. Complete keyword lists and formulas appear in Table S1 (columns P–AA).

C2 (data availability) scored 1 if URLs, DOIs, or repository names (Zenodo, GitHub, Figshare, MetaboLights, GNPS, NCBI) appeared. C3 (code availability) scored 1 if repository terms (GitHub, GitLab, PyPI, CRAN, Bioconductor) were present, and 0 if ``not available'' appeared. C4 (standardised I/O) scored 1 if recognised formats appeared in both input and output fields (mzML, mzTab, SMILES, InChI, CSV, JSON, among others). C5 scored 1 if database names (ChEMBL, PubChem, HMDB, KEGG, GNPS, MassBank, among others) appeared. C6 scored 1 if workflow engines or containers (Nextflow, Snakemake, Galaxy, Docker, Conda) appeared. Table S1 contains platform metadata, search strings, keyword lists, and scoring formulas (columns P–AA: 12 columns containing input/output formats, database terms, and scoring formulas for C2–C6).

A single author performed all coding to ensure uniform criterion application. Because scoring relied on explicit documentary evidence---a URL either appeared or it did not; a keyword was either present or absent---interpretive judgement was minimal. The complete coding matrix (Table S2) is publicly available for independent verification. 

\subsection{Statistical analysis}

All analyses used Microsoft Excel for Microsoft 365 (Version 2509, Build 16.0.19231.20246, 64-bit, locale en-GB, Windows 11). We used Pearson's chi-square tests to assess domain differences in pillar adoption. Where expected cell counts fell below 5, Fisher's exact test confirmed results (conclusions unchanged). Cramér's V quantified effect size: negligible (<0.1), small (0.1–0.3), medium (0.3–0.5), or large (>0.5), following Cohen (1988). Phi coefficients quantified pairwise pillar associations; odds ratios with 95\% CI derived from 2$\times$2 tables. Bonferroni correction controlled multiple testing ($\alpha$ = 0.0083 for 6 domain comparisons; $\alpha$ = 0.0033 for 15 pairwise associations), chosen for its conservatism given the exploratory nature of the analysis.  Results appear in Tables S7–S9.

\paragraph{Temporal stratification.} To assess whether reproducibility practices evolved over time, we stratified tools into three publication periods: 2004--2015 ($n=19$), 2016--2019 ($n=23$), and 2020--2025 ($n=61$). We chose period boundaries to reflect policy milestones rather than equal durations: 2004–2015 represents the pre-mandate era, 2016–2019 the transition period during which major journals adopted data-sharing requirements, and 2020–2025 the post-mandate era when such policies became near-universal in high-impact journals. We operationalised openness as the means of C2, C3 and C5; operability as the means of C1 and C6.

\subsection{Limitations and data availability}

We sampled for coverage, which may miss tools with limited documentation. Binary scoring does not capture gradations. Single-coder assessment ensured consistent criterion application but precluded inter-rater reliability calculation. Temporal bias may affect C6 scores: containerisation technologies emerged after 2013, retrospectively disadvantaging earlier tools.

Supplementary tables are deposited on GitHub

(https://github.com/Sarah-fheed/NTA-Platform-Audit-Supplementary-Data): Table S1 (103 tools with 35 metadata columns including pillar scores and Excel formulas), Table S2 (binary coding matrix: 103 tools × 6 pillars), Table S3 (domain distribution), Table S4 (overall adoption rates), Table S5 (15 pillar co-occurrence pairs), Table S6 (domain-specific adoption rates), Table S7 (phi coefficients with p-values, odds ratios, and 95\% CI for 15 pairwise associations), Table S8 (phi coefficient correlation matrix), Table S9 (chi-square domain comparison tests with Cramér's V). Source publications are archived on Zenodo (https://doi.org/10.5281/zenodo.17715417).

\section{Results}\label{sec:results}

We assessed 103 LC/GC--HRMS NTA tools published between 2004 and 2025 against six reproducibility pillars (C1–C6).

\subsection{Corpus composition}

\begin{figure}[htbp]
\centering
\includegraphics[width=0.85\textwidth]{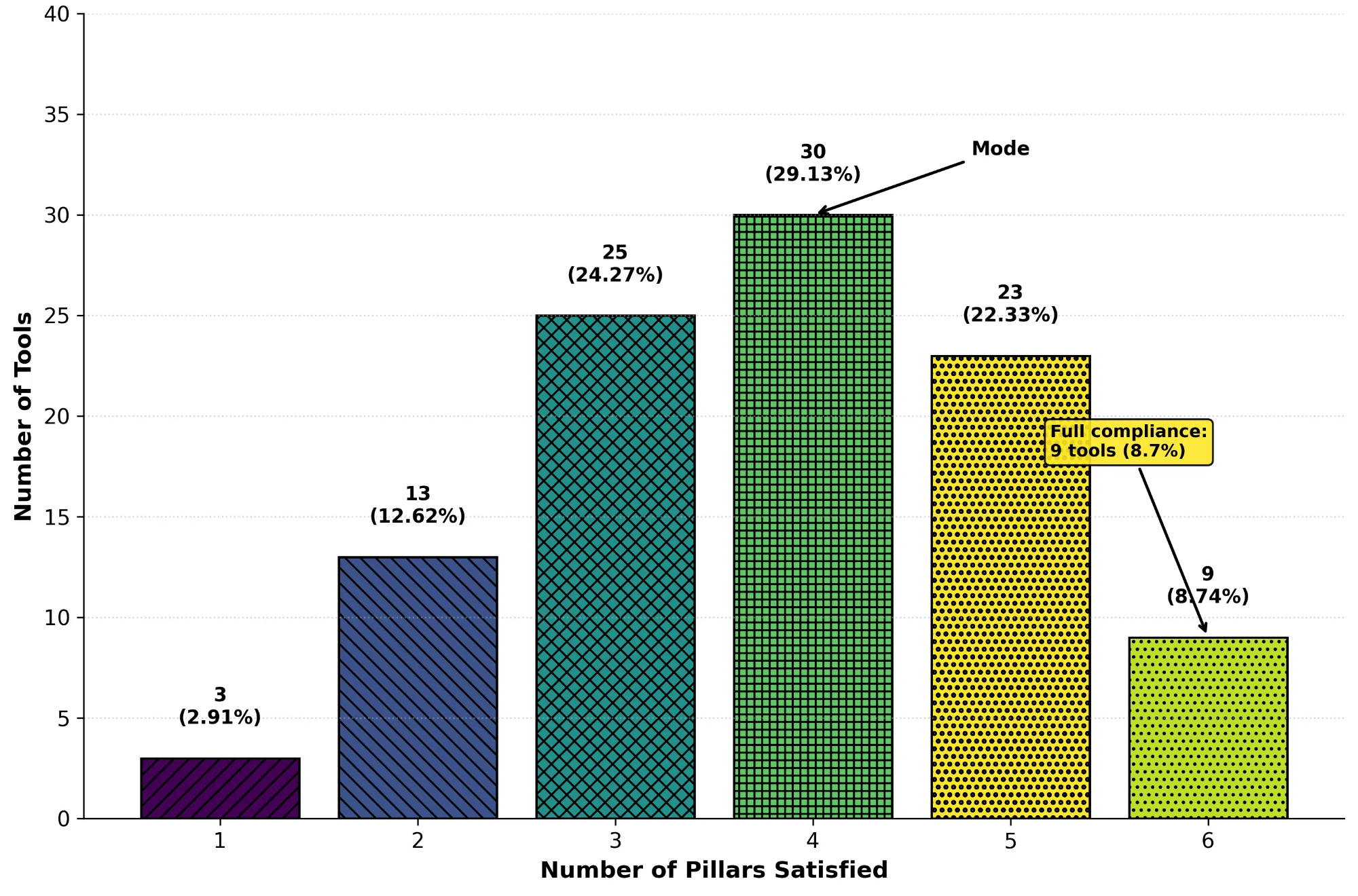}
\caption{Distribution of pillar counts across 103 NTA tools. The mode 
was four pillars (30 tools, 29.13\%); full compliance (six pillars) 
was observed in only 9 tools (8.74\%). Bars use viridis colour scale with 
distinct hatching patterns for print accessibility. Raw counts derive 
from the binary coding matrix (Table~S2).}
\label{fig:pillar-distribution}
\end{figure}

Health comprised the largest domain (51/103, 49.51\%), followed by Pharma (31/103, 30.10\%) and Chemistry (21/103, 20.39\%). Pillar compliance varied widely. Nine tools satisfied all six pillars (8.74\%); three satisfied only one (2.91\%). The most frequent count was four pillars, observed in 
30 tools (29.13\%), followed by three (25, 24.27\%), five (23, 22.33\%), and two (13, 12.62\%). Figure~\ref{fig:pillar-distribution} displays this distribution, showing that full compliance remains uncommon. 

Domain-specific pillar counts are detailed in Table~\ref{tab:domain-pillar}.

\subsection{Overall pillar adoption}

\begin{figure}[htbp]
\centering
\includegraphics[width=0.85\textwidth]{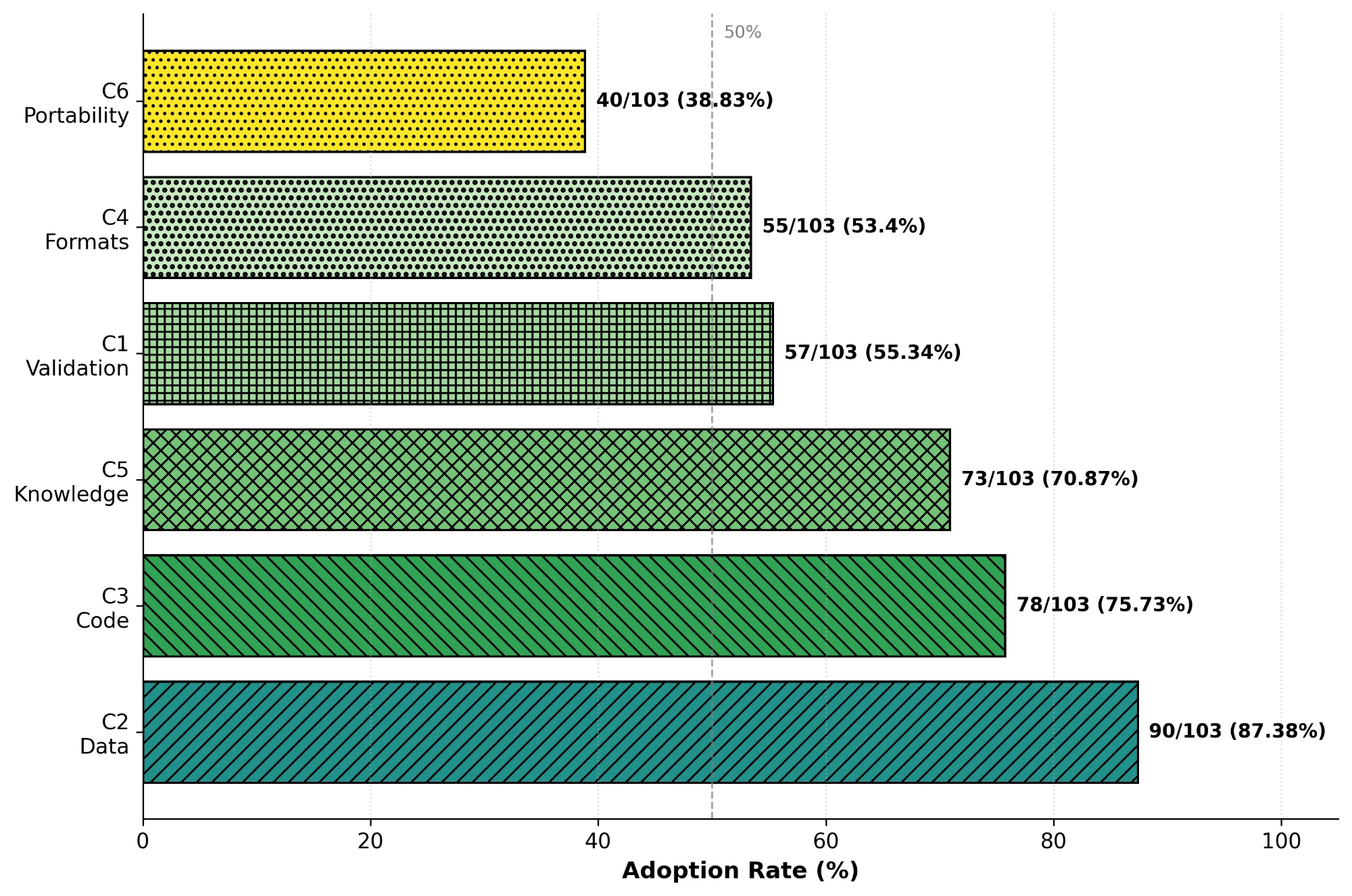}
\caption{Overall pillar adoption rates across 103 LC/GC--HRMS NTA tools, 
sorted by adoption percentage. Data availability (C2) reached 87.38\%; 
portable implementation (C6) remained lowest at 38.83\%. Bars are coloured 
using a viridis gradient with hatching patterns for print accessibility. 
Raw counts and percentages derive from Table~S4.}
\label{fig:overall-adoption}
\end{figure}

Data availability reached the highest rate (C2, 90/103, 87.38\%), followed by code availability (C3, 78/103, 75.73\%) and knowledge integration (C5, 73/103, 70.87\%). Laboratory validation (C1, 57/103, 55.34\%) and standardised I/O (C4, 55/103, 53.40\%) showed moderate adoption. Portable implementation remained the least common pillar 
(C6, 40/103, 38.83\%). Figure~\ref{fig:overall-adoption} visualises this gradient from openness to portability. Complete adoption rates appear in Table S4.

\begin{table}[ht]
\centering
\caption{Pillar adoption by domain. Counts (percentages) of tools satisfying each pillar.}
\label{tab:domain-pillar}
\begin{tabular}{lcccccc}
\toprule
Domain & C1 & C2 & C3 & C4 & C5 & C6 \\
\midrule
Health (n=51) & 30 (58.82) & 43 (84.31) & 37 (72.55) & 22 (43.14) & 32 (62.75) & 22 (43.14) \\
Pharma (n=31) & 15 (48.39) & 27 (87.10) & 24 (77.42) & 20 (64.52) & 24 (77.42) & 8 (25.81) \\
Chemistry (n=21) & 12 (57.14) & 20 (95.24) & 17 (80.95) & 13 (61.90) & 17 (80.95) & 10 (47.62) \\
\midrule
Total (N=103) & 57 (55.34) & 90 (87.38) & 78 (75.73) & 55 (53.40) & 73 (70.87) & 40 (38.83) \\
\bottomrule
\end{tabular}
\end{table}

\subsection{Temporal evolution of pillar adoption}

Reproducibility practices shifted substantially over the 21-year observation period (Table~\ref{tab:temporal}). Openness pillars---data availability (C2), code availability (C3) and database integration (C5)---rose from a combined mean of 56.1\% in 2004--2015 to 86.3\% in 2020--2025, a gain of 30.2 percentage points. Code availability showed the largest increase, rising from 42.1\% (8/19) in the earliest period to 85.2\% (52/61) in the most recent, a difference of 43.1 percentage points.

\begin{table}[htbp]
\centering
\caption{Pillar adoption rates (\%) by publication period. Openness = mean of C2, C3, C5; Operability = mean of C1, C6. Gap = Openness $-$ Operability.}
\label{tab:temporal}
\begin{tabular}{@{}lccccccccc@{}}
\toprule
Period & $n$ & C1 & C2 & C3 & C4 & C5 & C6 & Openness & Operability \\
\midrule
2004--2015 & 19 & 68.4 & 73.7 & 42.1 & 47.4 & 52.6 & 42.1 & 56.1 & 55.3 \\
2016--2019 & 23 & 56.5 & 82.6 & 78.3 & 43.5 & 60.9 & 47.8 & 73.9 & 52.2 \\
2020--2025 & 61 & 50.8 & 93.4 & 85.2 & 59.0 & 80.3 & 34.4 & 86.3 & 42.6 \\
\midrule
$\Delta$ (Early$\to$Recent) & --- & $-$17.6 & +19.7 & +43.1 & +11.6 & +27.7 & $-$7.7 & +30.2 & $-$12.6 \\
\bottomrule
\end{tabular}
\end{table}

Operability pillars moved in the opposite direction. Validation (C1) declined from 68.4\% (13/19) to 50.8\% (31/61), a reduction of 17.6 percentage points. Portable implementation (C6) remained low throughout, fluctuating between 34.4\% and 47.8\% with no clear trend. The combined operability mean fell from 55.3\% to 42.6\%.

The resulting gap between openness and operability widened from 0.9 percentage points in 2004--2015 to 43.7 percentage points in 2020--2025. Despite improved sharing, full compliance (6/6 pillars) remained essentially flat: 10.5\% (2/19) in the earliest period, 8.7\% (2/23) in the middle period, and 8.2\% (5/61) in the most recent. We interpret this pattern as evidence that journal data-sharing mandates successfully increased availability without correspondingly improving executability.
This finding reframes the reproducibility challenge. The problem is not that NTA tools lack openness---most now share data and code. The problem is that openness advanced while operability stagnated, creating a structural asymmetry that policy interventions have yet to address. The implications for underserved domains are examined in Section~\ref{sec:discussion}.

\subsection{Adoption by domain}

\begin{figure}[htbp]
\centering
\includegraphics[width=0.95\textwidth]{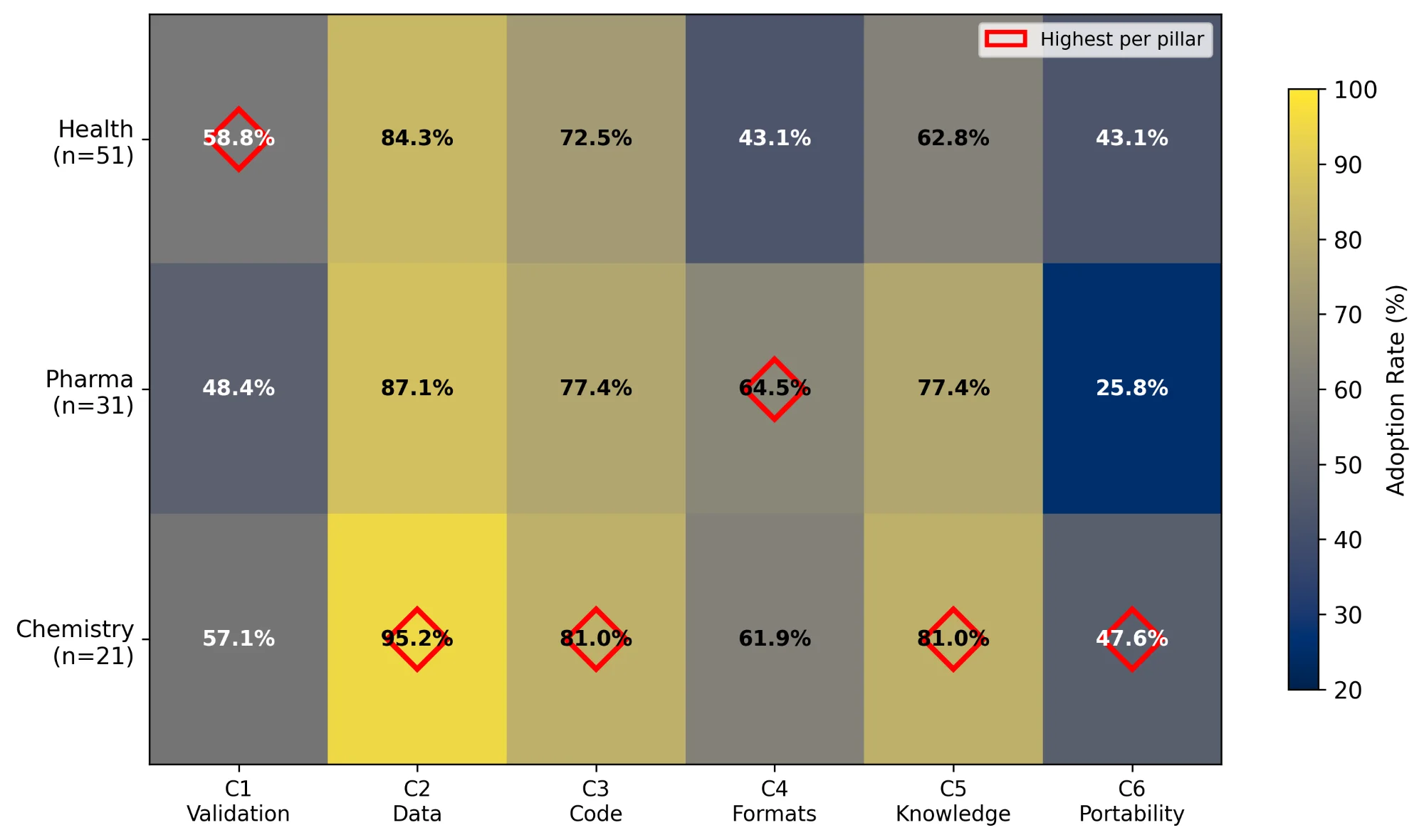}
\caption{Domain-wise adoption of reproducibility pillars ($N=103$). Each 
cell reports the percentage of tools within a given domain (Health, Pharma, 
Chemistry) satisfying each pillar (C1--C6). Diamonds indicate the highest 
adoption rate for each pillar. Chemistry led four of six pillars; portable 
implementation (C6) remained below 50\% across all domains. Cividis colormap 
ensures colorblind accessibility. Percentages derive from Table~S6; raw 
counts appear in Table~\ref{tab:domain-pillar}.}
\label{fig:domain-heatmap}
\end{figure}

Chemistry showed the highest adoption in four of six pillars: (C2, 95.24\%), (C3, 80.95\%), (C5, 80.95\%), and (C6, 47.62\%). Health showed the highest rate for (C1, 58.82\%), while Pharma showed the highest for (C4, 64.52\%). Portable implementation (C6) remained below 50\% in all three domains. (Figure~\ref{fig:domain-heatmap}) maps these patterns, with diamonds marking the leading domain for each pillar. Domain-specific percentages appear in Table S6.

Chi-square tests examined whether pillar adoption differed across domains. None reached significance after Bonferroni correction ($\alpha$ = 0.0083). The largest difference appeared for C4 ($\chi^2$ = 4.31, $p$ = 0.116, Cramér's $V$ = 0.14); effect sizes ranged from negligible to small (Cramér's $V$ = 0.06--0.14). Complete results appear in Table~S9.

\subsection{Co-occurrence patterns}

\begin{figure}[htbp]
\centering
\includegraphics[width=0.75\textwidth]{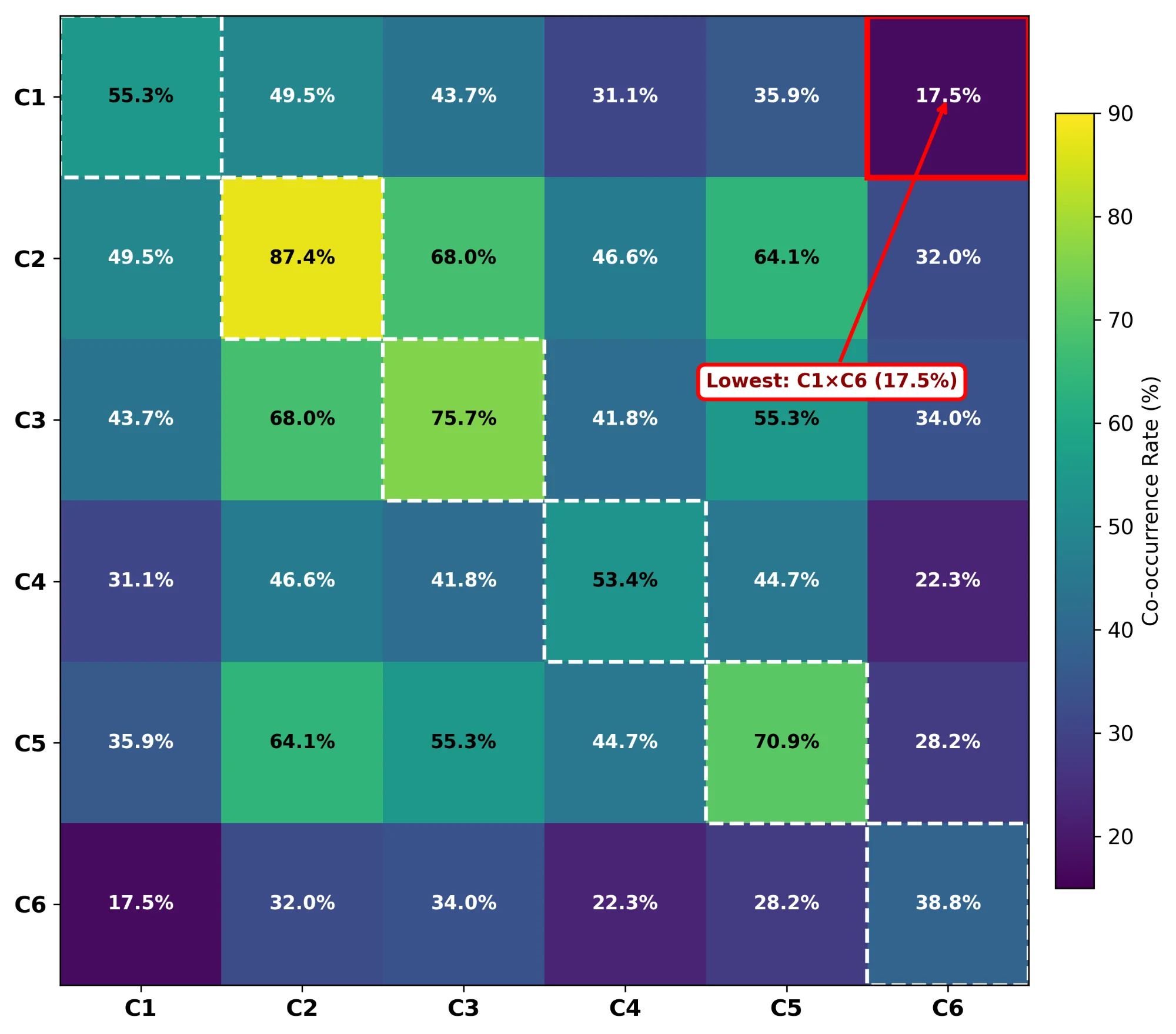}
\caption{Pillar co-occurrence matrix ($N=103$). Diagonal cells (dashed 
borders) show individual adoption rates; off-diagonal cells show pairwise 
co-occurrence percentages. (C1$\times$C6, 17.48\%, thick black border) represents 
the lowest co-occurrence---validated tools were least likely to support 
portable implementation. Viridis colormap ensures colorblind accessibility. 
Complete data appear in Table~S5.}
\label{fig:cooccurrence}
\end{figure}

The three highest co-occurrence rates were (C2$\times$C3) tools satisfying both pillars =(70/103, 67.96\%), (C2$\times$C5, 66/103, 64.08\%), and (C3$\times$C5, 57/103, 55.34\%). The three lowest were (C1$\times$C6, 18/103, 17.48\%), (C4$\times$C6, 23/103, 22.33\%), and (C5$\times$C6, 29/103, 28.16\%). Figure~\ref{fig:cooccurrence} displays these patterns as a symmetric matrix; the (C1$\times$C6) cell is highlighted as the lowest co-occurrence in the corpus.

Complete pairwise frequencies appear in Table S5.

\subsection{Pairwise associations}

Table~\ref{tab:phi-selected} presents phi coefficients and odds ratios 
for selected pillar pairs. (C4$\times$C5, Standardised I/O $\times$ 
Knowledge Integration) showed the strongest association ($\phi$ = 0.301, 
OR = 3.98, 95\% CI: 1.59--9.92, $p$ = 0.0023), indicating tools supporting 
standardised formats were nearly four times more likely to integrate 
external databases. (C3$\times$C6, Code $\times$ Portability) showed the 
second strongest association ($\phi$ = 0.219, OR = 3.26, 95\% CI: 
1.11--9.56, $p$ = 0.0264).

\begin{figure}[htbp]
\centering
\includegraphics[width=0.95\textwidth]{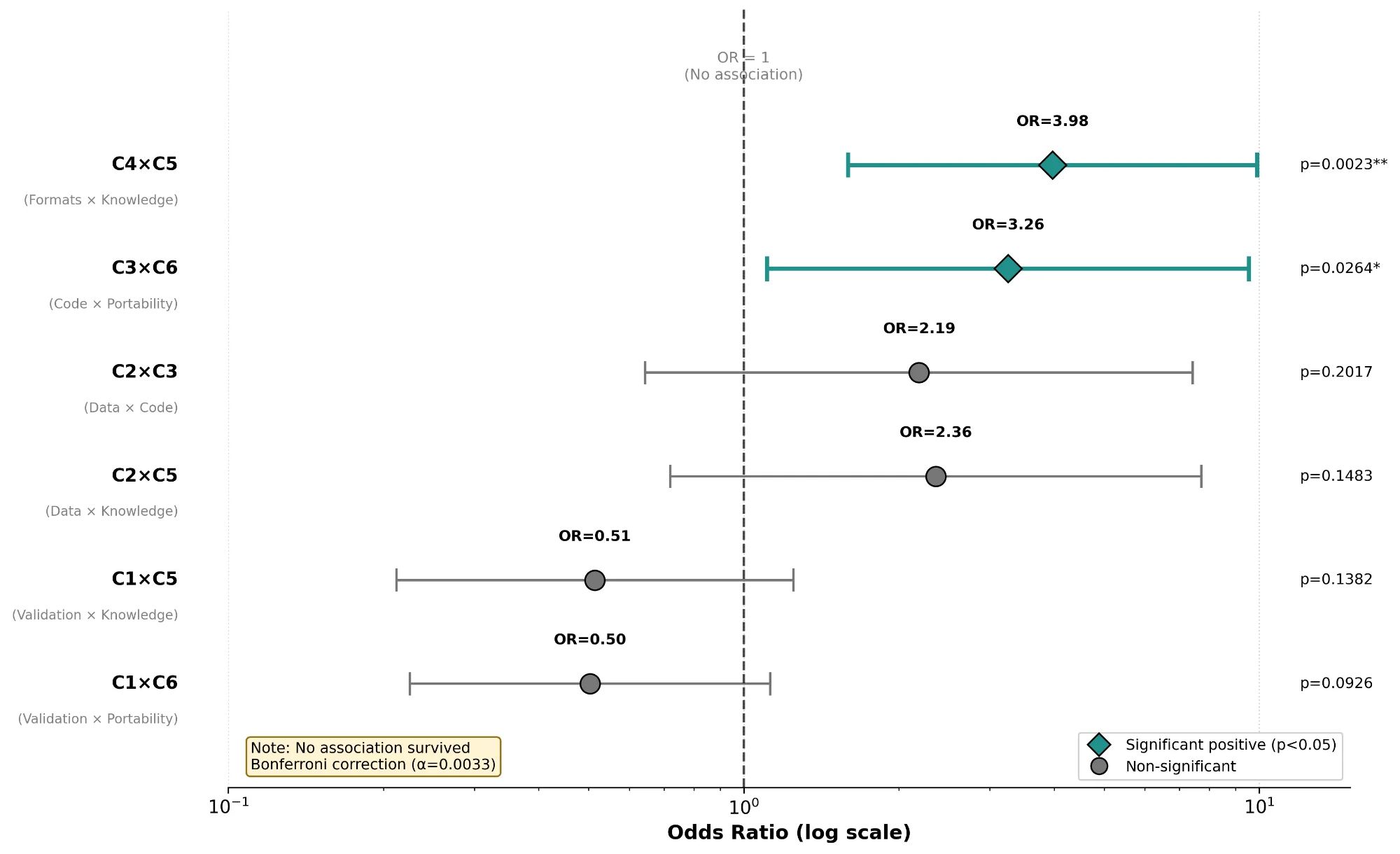}
\caption{Odds ratios with 95\% confidence intervals for selected pillar 
pairs. (C4$\times$C5 showed the strongest positive association, OR = 3.98, 
$p$ = 0.0023); (C1$\times$C6 showed a negative trend, OR = 0.50, $p$ = 0.093). 
Diamonds indicate pairs reaching uncorrected significance ($p$ < 0.05); 
circles indicate non-significant pairs. The dashed line marks OR = 1 
(no association). No pair remained significant after Bonferroni correction 
($\alpha$ = 0.0033). Data derive from Table~S7.}
\label{fig:forest-plot}
\end{figure}

Two pairs showed negative point estimates:  (C1$\times$C6, $\phi$ = --0.166, OR = 0.50, 95\% CI: 0.23--1.13) and (C1$\times$C5, $\phi$ = --0.146, OR = 0.51, 95\% CI: 0.21--1.25), suggesting a possible negative trend between validation and portability, though confidence intervals include unity and neither reached significance. Figure~\ref{fig:forest-plot} visualises these associations as a forest plot.

After Bonferroni correction ($\alpha$ = 0.0033), no association remained significant. Full results appear in Table S7.

\begin{table}[ht]
\centering
\caption{Phi coefficients and odds ratios for selected pillar pairs.}
\label{tab:phi-selected}
\begin{tabular}{lcccc}
\toprule
Pair & $\phi$ & $p$ & OR & 95\% CI \\
\midrule
C4$\times$C5 & 0.301 & 0.0023** & 3.98 & 1.59--9.92 \\
C3$\times$C6 & 0.219 & 0.0264* & 3.26 & 1.11--9.56 \\
C1$\times$C6 & $-$0.166 & 0.0926 & 0.50 & 0.23--1.13 \\
C1$\times$C5 & $-$0.146 & 0.1382 & 0.51 & 0.21--1.25 \\
C2$\times$C3 & 0.126 & 0.2017 & 2.19 & 0.64--7.43 \\
\bottomrule
\end{tabular}

\smallskip
\noindent\footnotesize{*$p$ < 0.05; **$p$ < 0.01; No pair significant after Bonferroni ($\alpha$=0.0033)}
\end{table}

Beyond pillar patterns, we examined domain coverage. No tool among 103 addressed food-safety applications; this gap is examined in the Discussion.

\section{Discussion}\label{sec:discussion}

\subsection{The openness-portability asymmetry}\label{sec:asymmetry}

By 2015, openness already exceeded operability. What changed since then is the acceleration: the gap widened from 0.9 to 43.7 percentage points. Nearly nine in ten tools now share their data; three in four share their code. Yet fewer than four in ten can be rerun outside their originating laboratory. This asymmetry is architectural, not incidental: it reflects what journals required (sharing) versus what they did not (execution). No pairwise association survived Bonferroni correction (Table~\ref{tab:phi-selected}). We interpret these findings as descriptive tendencies reflecting how NTA tools present themselves as reproducible objects, not as statistically confirmed dependencies.
The absence of significant associations itself suggests that pillars operate independently tools adopt openness practices without necessarily investing in portability infrastructure.

The temporal pattern observed in Section~3.3 reframes this asymmetry as a divergence rather than a static gap. In 2004--2015, openness and operability were nearly balanced (56\% vs 55\%). By 2020--2025, the gap had widened to 44 percentage points. Journal data-sharing mandates appear to have succeeded in one dimension while leaving another untouched.

The central question is not whether these tools work, but whether they can be trusted, examined and reused outside the laboratory that built them. For regulators and downstream decision-makers, this matters concretely. A chemical call from an NTA workflow can trigger product recalls, inform public-health advisories or justify enforcement actions, and may later need to withstand legal challenge \cite{hollender2023norman}. Under that standard, regulatory-grade reproducibility requires that an informed third party can rerun the workflow, with declared inputs and parameters, and reach a concordant conclusion without privileged access \cite{goodman2016reproducibility}. Open is not defensible. We use 'tool' as an umbrella term encompassing platforms, pipelines, packages, and workflows.

The data reveal a consistent asymmetry. Openness does not guarantee portability the high co-occurrence of (C2$\times$C3, 70/103, 67.96\%) masks a modest association ($\phi$ = 0.126) that did not survive correction. Tools share data and code without ensuring replayability elsewhere \cite{stodden2018enabling}.

\subsection{Why openness improved} \label{sec:openness_improve}

The rise in C2, C3 and C5 adoption almost certainly reflects journal data-sharing mandates introduced since approximately 2015, together with the maturation of repository infrastructure. These policies are enforceable at submission: reviewers and editors can verify that a Data Availability Statement exists and that a repository link resolves. Code availability (C3) showed the largest gain (+43.1~pp), consistent with policies that explicitly require code deposition. The pattern suggests that what journals mandate, developers provide.

\subsection{Why operability did not follow}\label{sec:Operability}

The decline in C1 and C6 is harder to enforce by policy. Validation (C1) requires experimental effort---reference standards, spike-recovery tests, cross-laboratory comparisons---that reviewers cannot easily verify from a manuscript. Portable implementation (C6) demands containerisation or workflow-engine integration that many developers lack time or incentive to provide. The result is a policy asymmetry: current mandates address \emph{what} is shared but not \emph{whether it works}. The 17.6~pp decline in C1 warrants attention: newer tools are less likely to report validation than older tools, even as analytical complexity scales.

Containerisation emerged as a strong correlate of compliance. Tools providing Docker or Singularity images ($n=15$) achieved a mean score of 4.87 pillars versus 3.64 for non-containerised tools ($n=88$; Mann--Whitney $U=1017.5$, $p=0.0003$). Full compliance was seven times more common among containerised tools (33.3\% vs 4.5\%). We interpret containerisation less as a cause than as a marker of developer diligence, but the association suggests that promoting container adoption could accelerate C6 compliance.

\subsection{Domain-specific patterns}\label{sec:domain-patterns}

Each domain concentrates on pillars closest to its operational pressures.
Health-oriented tools showed the highest laboratory validation (C1, 58.82\%). This pattern aligns with clinical settings where analytical claims must withstand external reconstruction \cite{kiseleva2022transparency}. Pharmaceutical tools led standardised I/O (C4, 64.52\%), aligning with regulatory submission formats and pharmacovigilance workflows that demand interoperable data exchange \cite{bhirud2024nitrosamine}.

Chemistry tools led four of six pillars: data availability (C2, 95.24\%), code availability (C3, 80.95\%), knowledge integration (C5, 80.95\%), and portable implementation (C6, 47.62\%), reflecting established practices of open-source development and shared spectral repositories \cite{Heller2015}. These differences did not reach statistical significance after correction ($\chi^2$ = 0.64–4.31, all $p$ > 0.05, Bonferroni $\alpha$ = 0.0083), so we treat them as tendencies rather than confirmed domain effects.

Portable implementation remains the bottleneck. C6 was the least adopted pillar (40/103, 38.83\%) and stayed below 50\% in all three domains (Figure~\ref{fig:domain-heatmap}). The co-occurrence of validation with portability was notably low (C1$\times$C6, 18/103, 17.48\%), the weakest pairing in the corpus. The asymmetry is structural. In regulated settings, refactoring a validated tool into containers triggers re-validation—so teams avoid it \cite{colombo2020value,pamungkas2024architecture}. The result is strong internal assurance but limited external portability. Community workflow infrastructures such as Galaxy, Nextflow and Snakemake have demonstrated that containerised execution can stabilise behaviour and emit provenance as a first-class artefact \cite{afgan2018galaxy,di2017nextflow,molder2021sustainable}. The barrier is often procedural rather than computational: organisations optimise for continuity because architectural change incurs re-validation costs \cite{besker2018managing}.

The forest plot (Figure~\ref{fig:forest-plot}) quantifies this pattern. Tools adopting standardised formats (C4) were four times more likely to integrate curated databases (C5), reflecting shared design priorities 
in data interoperability. Code availability (C3) associated with portable implementation (C6), consistent with open-source development practices where containerisation accompanies public repositories.

\subsection{The food safety gap}\label{sec:food-gap}

These patterns reveal which domains are served and which are not. Health, Pharma and Chemistry are served (Table~\ref{tab:domain-pillar}). Food safety and environmental monitoring are not: no tool among 103 addressed food matrix contaminants or environmental pollutant screening.

\paragraph{Matrix complexity.} Food safety confronts 100+ matrix types—fruits, meats, grains, oils—each generating distinct interference signatures. A workflow validated on apple extract cannot assume transferability to olive oil without matrix-specific re-optimisation \cite{freitas2016, taylor2022}.

\paragraph{Library coverage.} Pharmaceutical tools target thousands of endogenous metabolites with authenticated reference spectra. Food safety requires coverage of >1000 registered pesticides (EU regulation alone), many lacking authenticated standards, plus transformation products from food processing \cite{menger2021, kruve2021}.

\paragraph{Regulatory burden.} FDA 21 CFR Part 11 requires audit trails for electronic records \cite{FDA21CFR11}. Cross-site reproducibility demands inter-laboratory validation—independent laboratories reaching equivalent conclusions under governed protocols.

The co-occurrence of validation and portability was lowest in the corpus (C1$\times$C6: 18/103, 17.48\%). Tools validated internally prove least portable externally—a tension food safety cannot accommodate.

Environmental screening faces comparable demands: PFAS and other emerging contaminants, heterogeneous matrices, cross-laboratory validation for regulatory acceptance \cite{barzen2017discovery,peter2018using}. The 2008 melamine incident \cite{pei2011}, Sudan dyes in spices \cite{rebane2010review}, nitrosamines in pharmaceuticals \cite{ruepp2021eu}—in each case, laboratories across jurisdictions had days to verify findings.

Chemistry tools offer a path forward. Among nine fully compliant tools (6/6 pillars), four originate from Chemistry, four from Health, one from Pharma. Their approaches—chromatographic alignment, spectral matching, compound annotation—transfer readily to food matrices. Adapting these may prove faster than building anew.

Melamine required verification across three continents within days. Today: 103 tools, none for food matrices, none combining validation, standardisation and portability. Food safety needs precisely (C1$\times$C4$\times$C6)—the combination current tools least provide.

\subsection{Relation to prior frameworks}\label{sec:prior-work}

Our work builds on prior efforts but addresses a different question. BP4NTA and FAIR prescribe what should be reported \cite{peter2021nontargeted,barker2022fair}; this audit examines what is actually delivered in externally inspectable form. We drew on both frameworks when operationalising our six pillars: (C2 and C3) reflect FAIR's accessibility and reusability expectations; (C4) aligns with interoperability; (C1) operationalises BP4NTA's emphasis on method validation. Yet both frameworks prescribe what should be reported; they do not audit whether current tools actually deliver these properties in externally inspectable form. Previous comparisons in metabolomics and exposomics focused on analytical performance, not architectural reproducibility \cite{hulleman2023critical,lennon2024harmonized}. The present audit asks a different question: not ``how well does this tool perform?'' but ``can an external party replay, audit and defend its outputs without privileged access?'' To our knowledge, no previous study has mapped NTA tools across multiple application domains against standardised reproducibility criteria.

\subsection{Limitations}\label{sec:limitations}

Several limitations apply. We scored only publicly inspectable behaviour; undisclosed internal controls were not inferred, so tools with strong internal governance but limited disclosure may be under-represented. 
Pillars were coded as binary signals with equal weight, preserving comparability but not capturing gradations. A single coder performed all assessments, ensuring consistency but precluding inter-rater reliability estimation. Because scoring relied on explicit documentary evidence---a URL either appeared or it did not; a keyword was either present or absent---interpretive judgement was minimal and inter-rater disagreement unlikely. The complete coding matrix (Table~S2) is provided for independent verification.
Binary scoring sacrifices granularity: a tool with minimal validation and one with extensive multi-site trials both score C1=1. Future audits could explore ordinal extensions that capture such gradations.

Survivorship bias may inflate early-period validation rates: tools from 2004--2015 that lacked validation may have been discontinued or under-documented, leaving validated tools over-represented in retrospective sampling.

English-language prioritisation may have excluded tools documented primarily in other languages, potentially under-representing contributions from non-Anglophone research communities.

The audit characterises infrastructure readiness, not analytical performance. Coverage-oriented sampling from NTA-focused sources does not guarantee exhaustive retrieval and may over-represent well-documented tools. The corpus nonetheless spans 21 years 
(2004--2025) and three established domains, supporting pattern identification within the surveyed landscape.

\subsection{Implications}\label{sec:implications}

The six pillars operationalised here may serve as a baseline for evaluating future tools addressing underserved domains. Strengthening (C1, C4 and C6) for food and environmental applications would shift NTA from documented to portable and externally defensible: a 
laboratory could replay another's workflow under governed conditions, and a regulator could audit the evidential chain from spectrum to annotation. Standardised formats such as mzML and mzTab \cite{Martens2011,Griss2014}, persistent identifiers such as InChIKey 
\cite{Heller2015}, and provenance schemas such as W3C PROV and RO-Crate \cite{belhajjame2013prov,SoilandReyes2022ROCrate} provide the technical substrate; the gap lies in adoption, not availability.

The audit reveals where NTA infrastructure stands; the path forward requires translating architectural openness into operational portability.Regulators gain procurement criteria. Developers learn where to focus limited resources. The broader community gets an imperfect but measurable baseline for evaluating future tools, including food-safety platforms. The gap is now quantified; what remains is to close it.

\section{Conclusions}\label{sec:conclusions}

The field learned to share. It did not learn to run. Between 2004 and 2025, openness climbed 30 points while operability dropped 12. Journals asked for repositories; they got repositories. They did not ask for execution—and did not get it. The gap widened from 0.9 to 43.7 percentage points---not because portability declined in absolute terms, but because openness raced ahead while operability stagnated.

This divergence is architectural, not incidental. Journal policies successfully mandated sharing; they did not mandate functionality. The result is a growing corpus of tools that are findable, accessible and transparent---yet difficult to execute outside their originating laboratories.

The findings reveal consistent asymmetries. Food safety remains unserved: no tool among 103 addresses food matrix contaminant screening. Openness does not ensure portability most tools share data and code, yet fewer than four in ten support containerised execution. Domain patterns reflect operational pressures: chemistry leads openness and transferability; pharma leads standardised formats; health leads validation.

The six-pillar framework offers a reusable benchmark for future development. Tools designed for food safety and similar emerging domains would benefit from prioritising laboratory validation (C1), standardised I/O (C4), and portable implementation (C6). Platform developers have a measurable target; decision-makers have criteria for evaluation; regulators have a framework for specifying architectural requirements.

Limitations single-coder assessment, binary scoring, English-language prioritisation are detailed in the Methods and Discussion. The complete platform×pillar matrix is provided in Table S2.

Three directions merit future work. First, do higher pillar scores predict better cross-laboratory agreement? A multi-site ring trial could test this. Second, can this framework guide development of food-safety-specific NTA platforms? A proof-of-concept addressing (C1, C4, and C6) for pesticide screening would test framework utility. Third, does (C6) adoption increase over time? Repeating this audit in three to five years would answer that question. The architectural requirements for food-safety NTA are now defined: validated workflows (C1), standardised food-matrix formats (C4), and containerised cross-laboratory execution (C6). Strengthening these pillars would shift NTA outputs from documented to defensibly transferable enabling workflows that move across institutions and withstand regulatory scrutiny. The architectural requirements for regulatory-grade reproducibility are now defined. What remains is implementation.

\backmatter

\section{List of abbreviations}

AUC: Area Under the Curve; 
BP4NTA: Benchmarking and Publications for Non-Targeted Analysis; 
CI: Confidence Interval; 
C1--C6: Reproducibility Pillars 1--6; 
FAIR: Findable, Accessible, Interoperable, Reusable; 
FDA: Food and Drug Administration; 
GC: Gas Chromatography; 
GNPS: Global Natural Products Social Molecular Networking; 
HMDB: Human Metabolome Database; 
HRMS: High-Resolution Mass Spectrometry; 
InChI: International Chemical Identifier; 
KEGG: Kyoto Encyclopedia of Genes and Genomes; 
LC: Liquid Chromatography; 
mzML: Mass Spectrometry Markup Language; 
mzTab: Mass Spectrometry Result File Format; 
NTA: Non-Targeted Analysis; 
OR: Odds Ratio; 
PFAS: Per- and Polyfluoroalkyl Substances; 
PROV: Provenance Data Model; 
RO-Crate: Research Object Crate; 
SMILES: Simplified Molecular Input Line Entry System

\section{Declarations}

\subsection{Availability of data and materials}

All data underlying this article are openly available.
Supplementary Tables: Tables S1–S9 provide platform metadata, binary coding 
matrices, domain distribution, adoption rates, co-occurrence analysis, and 
statistical tests.
Platform documentation: 103 PDF files archived at Zenodo 
(https://doi.org/10.5281/zenodo.17715417).
Analysis workbooks:
- Project name: NTA-Platform-Audit
- Project home page: https://github.com/Sarah-fheed/NTA-Platform-Audit-Supplementary-Data
- Archived version: https://doi.org/10.5281/zenodo.17715417
- Operating system(s): Platform independent
- Programming language: Microsoft Excel (VBA formulas)
- Other requirements: Microsoft Excel 2016 or later
- License: CC BY 4.0
- Any restrictions to use by non-academics: None

\subsection{Competing interests}
The authors declare no competing interests.

\subsection{Funding}
This work was supported by King Abdullah University of Science and Technology (KAUST) Office of Research Administration under Award Nos. REI/1/5234-01-01, REI/1/5414-01-01, REI/1/5289-01-01, REI/1/5404-01-01, REI/1/5992-01-01, URF/1/4663-01-01, Center of Excellence for Smart Health (KCSH) under award number 5932, and Center of Excellence on Generative AI under award number 5940.

\subsection{Authors' contributions}

SA conceived the study, designed the reproducibility framework, identified and assessed the tools, conducted statistical analyses, interpreted results, and drafted the manuscript.
SA2 contributed to manuscript revision.
XG supervised the project and revised the manuscript.
All authors read and approved of the final manuscript.

\subsection{Acknowledgements}
This publication is based on work supported by the King Abdullah University of Science and Technology (KAUST) Office of Research Administration (ORA) under Award Nos. REI/1/5234--01--01, REI/1/5414--01--01, REI/1/5289--01--01, REI/1/5404--01--01, REI/1/5992--01--01, URF/1/4663--01--01, Center of Excellence for Smart Health (KCSH), under award number 5932, and Center of Excellence on Generative AI, under award number 5940.

\subsection{Ethics approval and consent to participate}
Not applicable. This study analysed publicly available software 
documentation and involved no human or animal subjects.

\subsection{Consent for publication}
Not applicable.

\bibliography{References}
\end{document}